\documentclass[11pt,a4paper]{article}
\pdfoutput=1

\usepackage{jheppub}
\usepackage{color}
\usepackage{graphicx}
\usepackage{wrapfig,enumerate,slashed}
\usepackage[utf8]{inputenc}
\usepackage{adjustbox}

\usepackage{appendix}

\usepackage{footmisc}
\usepackage{amsmath}
\usepackage{wasysym} 
\usepackage{graphicx}
\usepackage{color}
\usepackage{comment}
\usepackage{hyperref}
\usepackage{subcaption}
\usepackage{slashed}
\usepackage{booktabs}

\newcommand{\rig}{\rightarrow}

\newcommand{\be}{\begin{eqnarray}}
\newcommand{\ee}{\end{eqnarray}}

\newcommand{\bee}{\begin{eqnarray}}
\newcommand{\eee}{\end{eqnarray}}
\newcommand{\beeq}{\begin{equation}}
\newcommand{\eeeq}{\end{equation}}

\usepackage{listings}
\usepackage{color}

\definecolor{dkgreen}{rgb}{0,0.6,0}
\definecolor{gray}{rgb}{0.5,0.5,0.5}
\definecolor{mauve}{rgb}{0.58,0,0.82}

\title{Unsupervised Event Classification with Graphs on Classical and Photonic Quantum Computers}

\author[a,b]{Andrew Blance}
\author[a]{and Michael Spannowsky}

\affiliation[a]{IPPP, Department of Physics, Durham University, Durham DH1 3LE, UK}
\affiliation[b]{Institute for Data Science, Durham University, Durham, DH1 3LE, UK}

\emailAdd{andrew.t.blance@durham.ac.uk}
\emailAdd{michael.spannowsky@durham.ac.uk}

\abstract{
Photonic Quantum Computers provides several benefits over the discrete qubit-based paradigm of quantum computing.  By using the power of continuous-variable computing we build an anomaly detection model to use on searches for New Physics. Our model uses Gaussian Boson Sampling, a $\#$P-hard problem and thus not efficiently accessible to classical devices. 
This is used to create feature vectors from graph data, a natural format for representing data of high-energy collision events. 
A simple K-means clustering algorithm is used to provide a baseline method of classification. We then present a novel method of anomaly detection, combining the use of Gaussian Boson Sampling and a quantum extension to K-means known as Q-means. This is found to give equivalent results compared to the classical clustering version while also reducing the $\mathcal{O}$ complexity, with respect to the sample's feature-vector length, from $\mathcal{O}(N)$ to $\mathcal{O}(\mbox{log}(N))$. Due to the speed of the sampling algorithm and the feasibility of near-term photonic quantum devices, anomaly detection at the trigger level can become practical in future LHC runs.
}

\begin{document}
\preprint{}
\maketitle
\flushbottom


\section{\label{Sec:Intro}Introduction}

The search for New Physics depends on our ability to separate Standard Model events from the much rarer and complex signal events from within the data from the LHC. It is desirable to perform data-driven searches, i.e. train on this data directly, without making specific assumptions on the new physics scenario realised by nature. While this allows the search to be widely applicable, it requires the classification method to learn and identify the important features of the background data to discriminate them from rare signal events which do not conform to the same features.
Within the realm of classical network methods, autoencoders have been designed for the purpose of anomaly detection and unsupervised event classification \cite{roy2020robust, Blance:2019ibf}. Other, cluster based methods have also been developed \cite{mikuni2020unsupervised}.

Careful consideration of how the data is represented can improve the performance of searches. Events in high energy physics naturally fit into graph structures. These structures allow us not only to define the features of individual constituents in the event but also how they are related to each other. Graphs have proven to be a powerful representation for LHC data \cite{Abdughani_2019, martinez2019pileup, Shlomi_2021} and combined with an anomaly detection method could be useful in data-driven searches. 

To further boost the performance of an anomaly detection search we propose the use of a quantum device. Quantum computing has shown to be able to solve classically hard problems, such as database searching and factoring \cite{Shor_1997, Grover:1996rk}. In HEP, quantum computers have become popular to solve a range of tasks. Annealers have been used to study field theories \cite{Abel:2020ebj, Abel:2020qzm} and optimisation problems \cite{cite-Higgs}. Quantum neural networks to solve classification problems have been built from quantum gates \cite{Blance:2020nhl}. Calculation of multi-particle interactions \cite{Jordan:2011ci,Garcia-Alvarez:2014uda,Jordan:2014tma,Jordan:2017lea,Preskill:2018fag,Moosavian:2019rxg,Alexandru:2019ozf, Alexandru:2019nsa, Lamm:2019uyc, Lamm:2020jwv} accomplished with the mapping of field theories onto quantum walks \cite{Marque-Martin:2018PRA,Arrighi:2018PRA,Jay:2019PRA,DiMolfetta:2020QIP} or a combination of quantum and classical ideas \cite{Lamm:2018siq,Harmalkar:2020mpd,Wei:2019rqy,Matchev:2020wwx} have also all been done using models built from quantum gates.

Previous work, mostly, focuses on quantum annealers or the discrete, qubit-based paradigm of quantum computing. While these models are very successful, another scheme can be employed. The continuous-variable (CV) model of quantum computer differs from its use of qumodes over qubits \cite{Lloyd_1999}. Data is embedded into these qumodes, an infinite-dimensional object. This is typically an electromagnetic field, allowing CV quantum devices to be constructed using quantum photonics hardware. Programming photonic quantum devices proceeds in stark similarity to qubit-based devices, with both allowing the construction of circuits from quantum gates. As qumodes are represented by an infinite-dimensional uncountable basis, the CV model is a more natural choice to simulate bosonic and continuous systems. In the particular interest of this paper is applications for graph data \cite{Bromley_2020} and machine learning \cite{Killoran_2019_NN}.

Photonic devices use single photon emissions, manipulated through squeezers and beamsplitters. A photonic system, even one that uses a non-interacting source of photons, can still exhibit quantum properties such as entanglement. An example of one such system is boson sampling \cite{aaronson2010computational}. Boson sampling techniques, such as Gaussian boson sampling (GBS), is an application of continuous-variable quantum computers where there is a clear advantage over classical devices \cite{Hamilton_2017}. A GBS device emits photons into an interferometer, generating a sample by counting the photons that exit the device. This setup can be constructed as a circuit in a quantum device. It is difficult to classically simulate the probability distribution of this process as it requires the calculation of the hafnian, a $\#$P problem. 

However, when access to GBS devices become more accessible samples can be generated at the rate of  $10^{5}$ every second \cite{Schuld_2020}. By embedding graph data into this device we can use it to generate a lower-dimensional representation of the data which is easier to handle when trying to build classifiers. These fast response times make GBS-based anomaly detection methods suitable to implement at the trigger level in future runs of the LHC. While the L1 trigger operates at $~1$ $\mu s$, the higher-level trigger functions at $< 100$ ms. A GBS device could produce around $100,000$ samples in a similar time frame, significantly more than is required in the method proposed here. 

The operation of a photonic device also presents some technical advantages over most qubit-based machines. Temperatures of less than 100 millikelvin are required for modern solid-state machines to operate efficiently \cite{hot_qubits}. Photonic devices are built from photonic hardware that can run at room temperature.

We aim to harness the power of continuous variable quantum computing, specifically Gaussian boson sampling, to survey particle physics events represented as graphs. These samples could then be used as input into various anomaly detection techniques. The anomaly detection method we use is built around the K-means clustering algorithm. We also present a quantum extension to this. Q-means clustering can be implemented on both a qubit-based, and qumode-based, quantum computer. In its most basic form, Q-means provides a substantial advantage over K-means. The size of the feature vector being used grows the time complexity of K-means linearly, whereas for Q-means it grows logarithmically, an exponential improvement \cite{lloyd2013quantum, kopczyk2018quantum}. 

Specifically, the process of anomaly detection we present has three steps:  (i) the creation of data and graphs, (ii) embedding of graphs in a lower dimension representation and (iii) the classification procedure. A classical method is shown as well as the use of quantum equivalents for parts (ii) and (iii). We apply our method to a search for hadronically decaying scalar resonances, produced through a Higgs-portal interaction in the $pp \rig HZ$ channel \cite{Falkowski:2010hi, Chen:2010wk}. To allow the process to trigger, we require the Higgs boson to recoil against a boosted leptonically-decaying Z boson. The major Standard Model background for such a signal is $pp \rig Z+\mathrm{jets}$.

These events can be represented as a set of graph adjacency matrices, weighted by the event constituent's features (ie. $m_{ij}$, $\Delta R$). We then find lower-dimensional embeddings for these objects and perform anomaly detection with them. To use as a baseline we create embeddings from the matrix eigenvalues and perform the classification using a K-means clustering algorithm. We then propose a novel method using GBS to create the embeddings and perform anomaly detection using a quantum equivalent of K-means known as Q-means. We find the GBS embedding method has improved performance compared to the classical version. 

The paper has the following structure: Section \ref{Sec:data} discusses the event generation and the process of encoding this into a graph structure. In Section \ref{Sec:classical} we introduce our first method of transforming the graph structure into something that can be used to train a classification algorithm. Here, we also present results as to how well the ``classical" method of vector embedding performs for anomaly detection. The photonic methodology is discussed in Section \ref{Sec:photonic}. We detail how a photonic circuit is created, how GBS sampling is performed and the results it gives when its samples are used for classification. Then, in Section \ref{Sec:qmeans} a quantum equivalent to K-means classification is adopted. Finally, a summary and conclusion is presented in Section \ref{Sec:conc}

\section{\label{Sec:data}Analysis Setup}
\subsection{Data Generation}

To use as our background and signal samples we generate $pp \rig Z+\mathrm{jets}$ and  $pp \rig HZ$ events, with subsequent decays $H \to A_1 A_2$, $A_2 \to gg$ and $A_1 \to gg$. All events have been generated with a centre of mass energy of $14$~TeV and a minimum $p_T$ of the hard process of at least $140$~GeV. We force the $Z$ boson to decay leptonically to either $e$ or $\mu$. We have set the Higgs mass to $125$~GeV, the $A_2$ mass to $40$~GeV and the mass of $A_1$ to be $60$~GeV. We use \textsc{Pythia 8.2} to generate events and perform parton showering \cite{Sj_strand_2015}. Such a scenario could be realised through derivative interactions between the Higgs boson and the pseudoscalars $A_1$ and $A_2$, which in turn form an effective, yet highly suppressed, interaction with gluons. Thus, their decay to gluons could still be prompt, whereas their direction production cross section in proton collisions was tiny. In such a scenario the observation of $A_1$ or $A_2$ would have to proceed through Higgs decays.

Our analysis follows the methodology used when using jet substructure to find the Higgs for our analysis \cite{Butterworth_2008, Soper_2010,Marzani:2019hun}. We will cluster our events into a fat jet, before reclustering its contents into ``microjets" \cite{Soper_2010,Soper:2012pb}. First, though, we impose a rapidity cut of 2.5 and a $p_T$ cut of $10$ ~GeV on all final state leptons. To reconstruct the Z boson we ensure to retain two charged leptons within $|y_l| \leq 2.5$ and require them to have an invariant mass of $80 \leq m_{ll} \leq 100$ GeV.  To explore the boosted region, we only consider events where the $p_T$ of the Z boson is greater than $150$~GeV. Based on the fat-jet properties only, signal events resemble background events very closely.

\begin{figure}[!t]
	\centering
	\begin{subfigure}{0.49\linewidth} \centering
		\includegraphics[width=\textwidth]{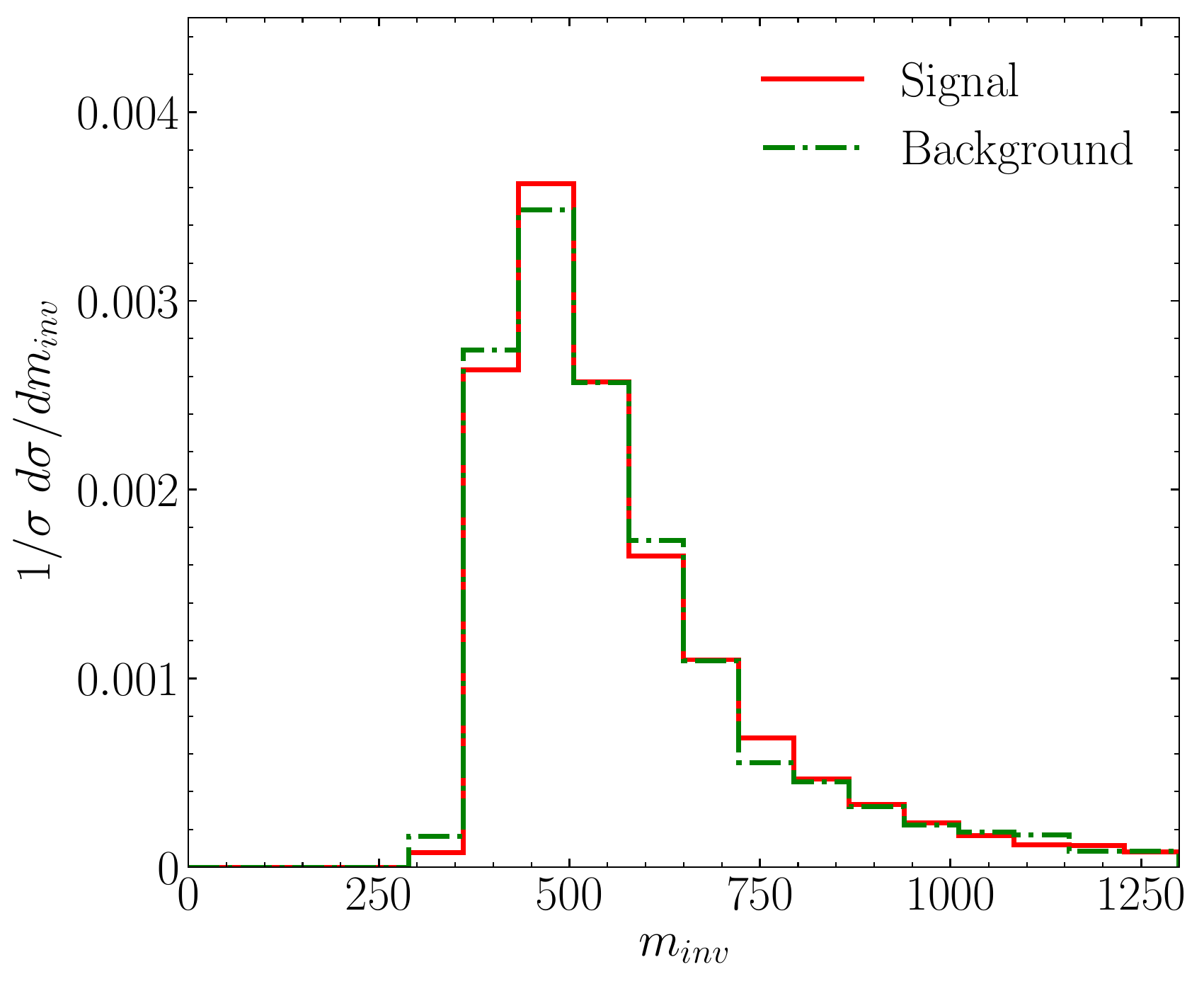}
		\caption{}
	\end{subfigure}
	\begin{subfigure}{0.49\linewidth} \centering
		\includegraphics[width=\textwidth]{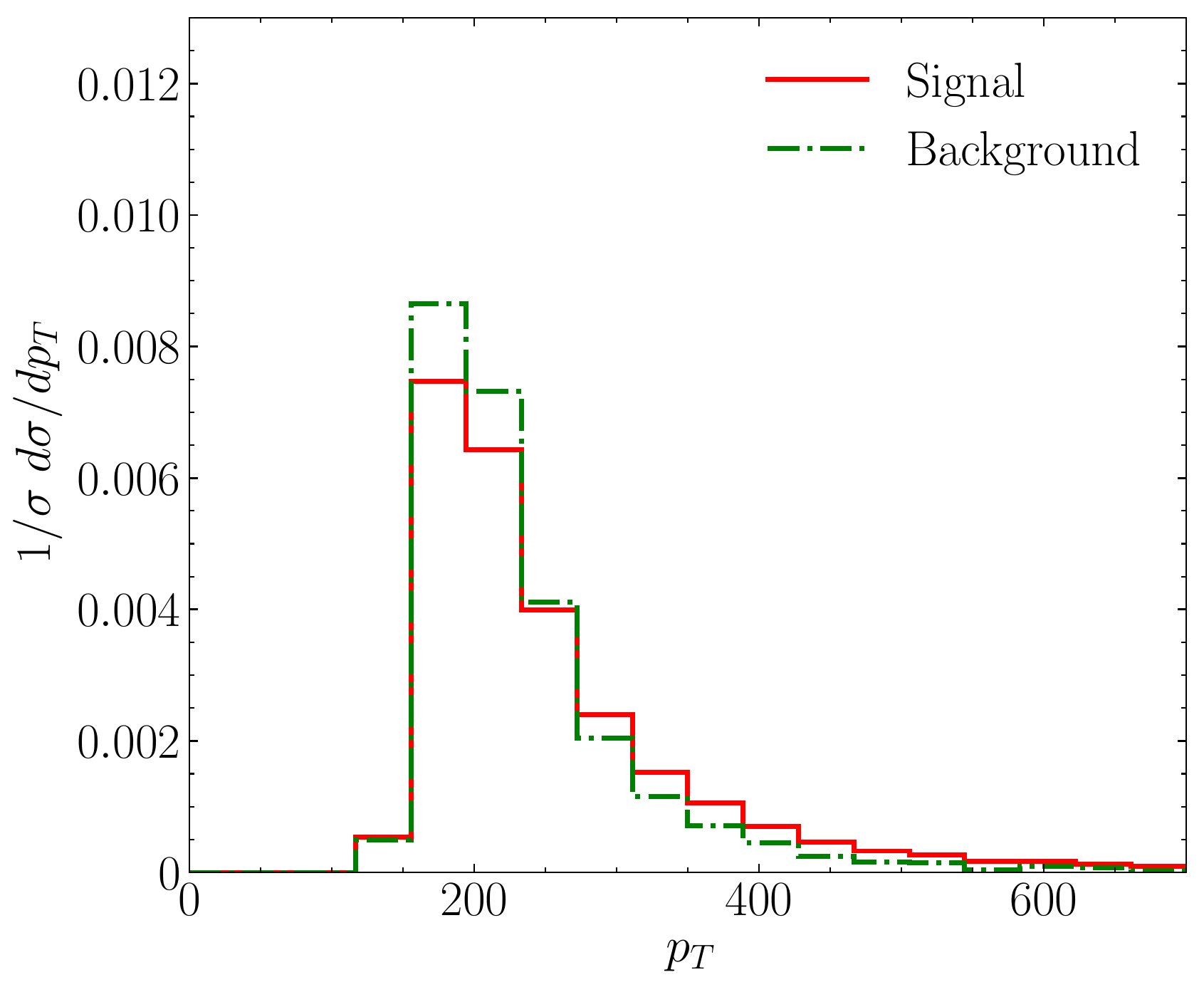}
		\caption{}
	\end{subfigure}
	\caption{(a) shows the invariant mass of the leptons and fat jet while (b) shows the transverse momentum of the fat jet.}
	\label{fig:first_clust}
\end{figure}

The remaining objects are subject to a rapidity cut of $5$. Using \textsc{FastJet} \cite{Cacciari_2012} we cluster these objects into jets using the Cambridge-Aachen algorithm with $R=1.5$ \cite{Dokshitzer_1997} and demand that there is at least one jet in the event with a transverse momentum of $p_T>150$~GeV. Figure \ref{fig:first_clust} shows the invariant mass of the leptons and fat jet in the event and also the transverse momentum of the fat jet.

\begin{figure}[!t]
	\centering
	\begin{subfigure}{0.49\linewidth} \centering
		\includegraphics[width=\textwidth]{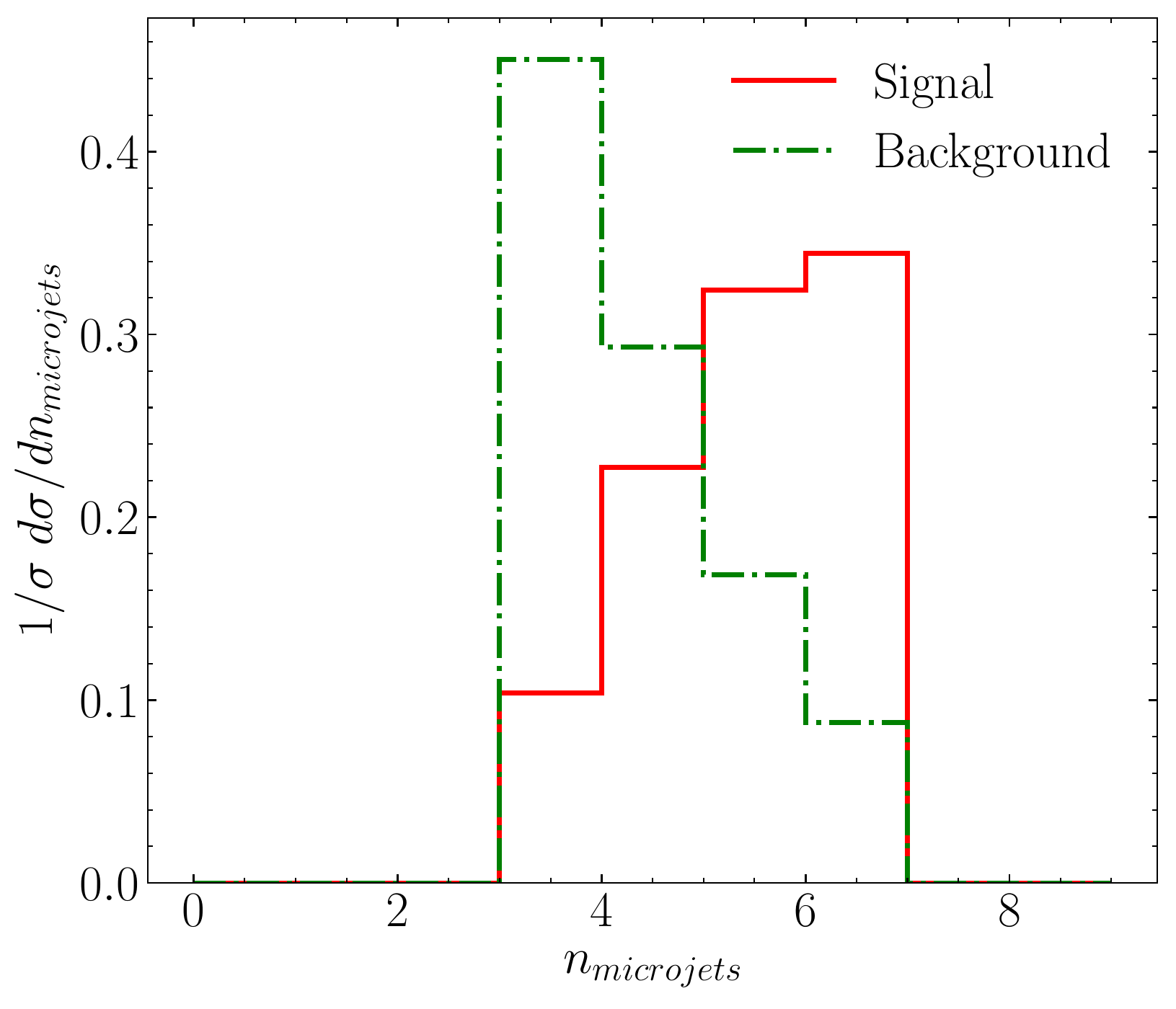}
		\caption{}
	\end{subfigure}
	\begin{subfigure}{0.49\linewidth} \centering
		\includegraphics[width=\textwidth]{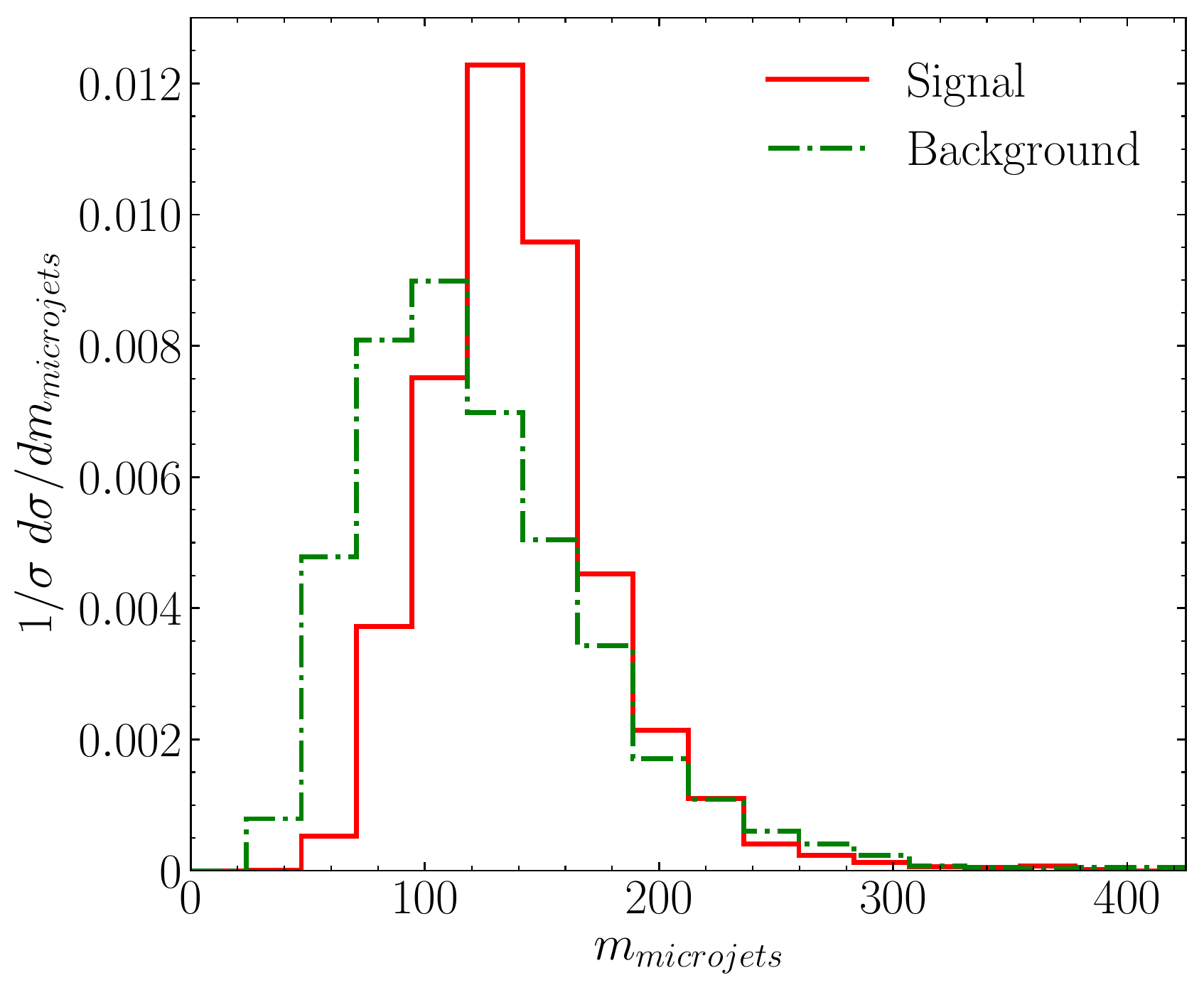}
		\caption{}
	\end{subfigure}
	\caption{(a) shows the number of microjets in each fat jet. (b) shows the total mass of each fat jet's collection of microjets.}
	\label{fig:second_clust}
\end{figure}

The hardest jet is then reclustered into a series of ``microjets" using the anti-kt algorithm \cite{Cacciari_2008}. Here, we choose $R=0.2$ and force a transverse momentum of $p_T>5$~GeV. Only events with 3-6 microjets are selected for analysis. The choice of this limit is to informed by the runtime of the quantum sampling process used. Larger graphs begin to take inhibitive amounts of time to be sampled on a simulator using our quantum sampling method. The number of jets chosen strikes a balance between the length of time it takes and being able to capture the maximum amount of information on the event. The number of microjets found, and the total mass of a fat jet's microjets is shown in Figure \ref{fig:second_clust}. These microjets are the objects used to construct our graphs. The result of the cuts we have applied is shown in Table \ref{Tab:cuts}. Here, we see the fraction of remaining events after each constraint is applied.

\begin{table}[]
	\centering
	\begin{tabular}{ c c c}
		\hline
		Cuts                                          & background   & signal   \\
		\hline \hline
		1 pair of leptons, $\eta<2.5$, $p_T > 10$ ~GeV & 0.74 & 0.54 \\
		\hline
		$p_{T_{ll}} > 150$ ~GeV,    $80$ ~GeV $<  M_{ll} < 100$ ~GeV                & 0.65 & 0.72 \\
		\hline
		1 fat jet with $p_T > 150$ ~GeV            & 0.94 & 0.97 \\
		\hline
		$2 <$ number of microjets $< 7$           & 0.42 & 0.62 \\
		\hline
		&       &      
	\end{tabular}
	\caption{Cut efficiencies relative to the previous event reconstruction step for signal and background events.}
	\label{Tab:cuts}
\end{table}

\subsection{Constructing the Graphs}

\begin{figure}[!t]
	\centering
	\begin{subfigure}{0.99\linewidth} \centering
		\includegraphics[width=\textwidth]{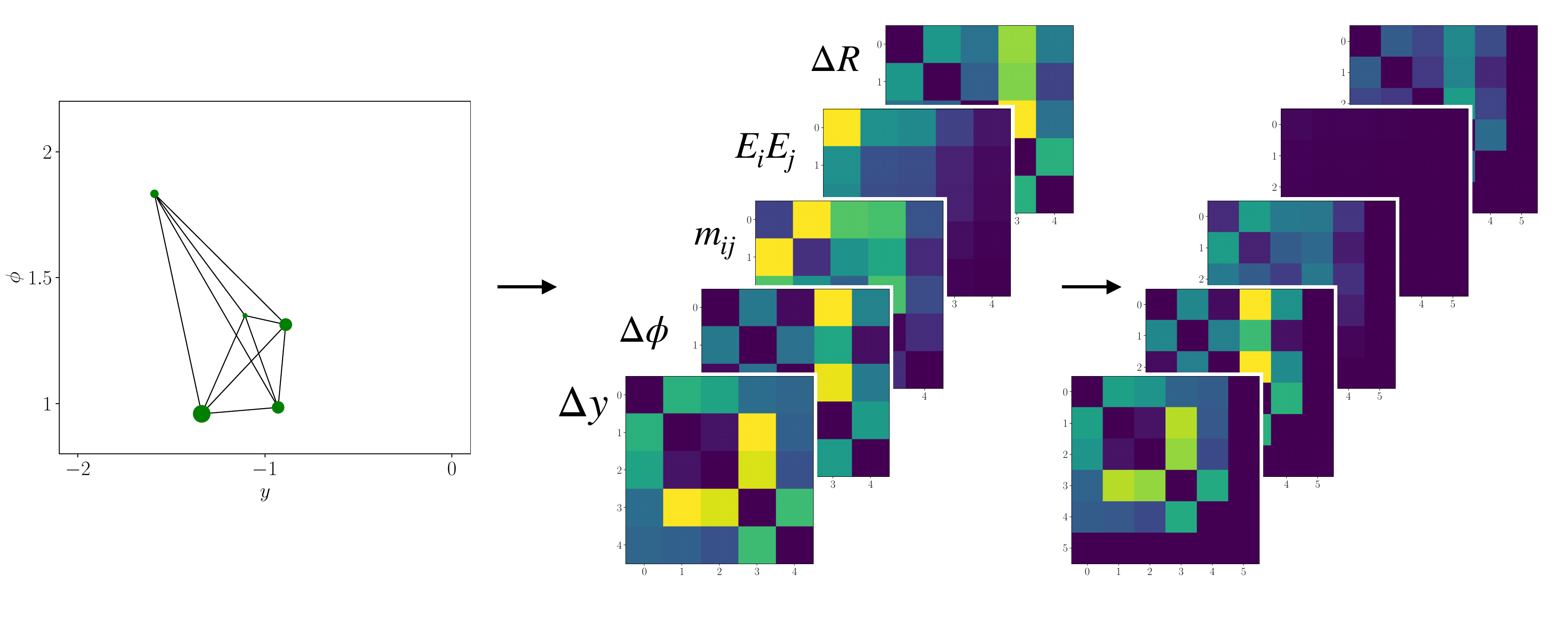}
	\end{subfigure}
	\caption{For a graph we find five weighted adjacency matrices from it, each one capturing features of the graph nodes. These five are then scaled and padded, giving the final matrices on the right. These can be embedded and used as input to classification algorithms.}
	\label{fig:matrix_maker}
\end{figure}

A method of  constructing graphs from our events is to use only the final states \cite{Abdughani_2019, martinez2019pileup}. Specifically, we use the microjets found from each event as the nodes to construct a set of graphs. We choose to connect nodes in the final state to every other node. Figure \ref{fig:matrix_maker} shows an example of a graph we may construct. Signal graphs in our sample contain on average 4.9 nodes, while background graphs contain 3.9. From here we can construct a set of  adjacency matrices that can be weighted to give detail on certain features of our event. Adjacency matrices are matrix representations of a graph where the rows and columns are defined by the graph nodes. Entries in the matrix are either 1 or 0, based on whether 2 nodes are connected. However, more information than this can be stored - the matrix entry can be weighted to show how strongly the nodes are connected. 

Some graph classification methods separate the graph information into feature and adjacency matrices \cite{kipf2017semisupervised}. 
In these cases, the adjacency matrix will include information on how the nodes in the graph are connected while the node's features are detailed in the feature matrix. However, for many graph sampling techniques, which require as input only symmetric matrices, a feature matrix cannot be readily included. We are limited to how we can manipulate adjacency matrices. Therefore, our aim is to construct a set of symmetric matrices that will include as much information about the node connections, and therefore our event, as possible. 
Our goal here does not require a model as complex as a neural network. We aim just to find a good representation of the graph, allowing us to use other methods to make classifications.

To begin with, an adjacency matrix can be constructed with connections between nodes weighted by their distance to each other in the $(y,\phi)$ plane. This distance is given by
\begin{equation}
\label{eq:graph_dist}
\Delta R = \sqrt{\Delta y^2 + \Delta \phi^2 },
\end{equation}
where $\Delta y$ is the difference in rapidity between particle $i$ and $j$, while $\Delta \phi$ is the difference between the azimuthal angle of particles $i$ and $j$. The dimensions of the resulting matrix will be determined by the number of nodes in the graph, whereas the matrix entries are found from Eq.~(\ref{eq:graph_dist}). We will eventually want to compare graphs, and hence matrices, of different sizes. Therefore, we pad each matrix with zeros on the right and lower sides such that they all have dimensions of $6\times6$. To be explicit, we can also construct weighted matrices from $\Delta \phi$ and $\Delta y$. Other adjacency matrices can be constructed using a measure of 2 objects energy $E_iE_j$ or invariant mass $m_{ij}$.

To be able to compare features of individual events with the global properties of the background event sample, each set of adjacency matrices are individually scaled with a constant
\begin{equation}
\label{eq:matrix_scaling}
s = 1/x,
\end{equation}
where $x$ is the maximum value in the set of matrices made from the background samples. For example, the $m_{ij}$ matrices are scaled by a term $s_{m_{ij}}$, while the $E_iE_j$ matrices are scaled by $s_{E_iE_j}$. Thus, the five matrices ($\Delta R$, $\Delta y$, $\Delta \phi$,  $E_iE_j$ and $m_{i,j}$) are all scaled accordingly.

Figure \ref{fig:matrix_maker} shows the process of taking one final event graph, finding a selection of weighted adjacency matrices ($\Delta R$, $\Delta y$, $\Delta \phi$,  $E_iE_j$ and $m_{i,j}$)  and then the final set of scaled and padded matrices. These adjacency matrices, like those in Figure \ref{fig:matrix_maker} will be embedded in a lower-dimensional space and then used as input into a classifier.

\section{\label{Sec:classical}Anomaly Detection on a Classical Computer}

To make predictions the graph data will be embedded into vectors. By being a 1-dimensional vector, rather than a 2-dimensional matrix, it is simpler to use them as input into a classifier. These vectors will then be used in an unsupervised K-means method of anomaly detection. What is being done here is somewhat different from kernel-based methods of graph classification. If one were to use the kernel framework the entire graph dataset would be embedded into a kernel. These kernels provide a measure of how similar each graph is to the other. This object could then be passed to a kernel-based classification algorithm (ie, a SVM) and from there predictions could be made. While this method is popular and has been shown to have predictive powers it is not without its limitations. By creating a graph-kernel one is required to use a kernel-based algorithm. However, by creating feature vectors from the graphs we can be more versatile when creating a model as the range of algorithms available to use is broader.

The adjacency matrices created in Section \ref{Sec:data} will be the objects used to create feature vectors. For the classical example we will simply take the eigenvalues of the five matrices. Each of the matrices gives us 6 eigenvalues. When these 5 objects are combined it results in a vector of length 30.  Finally, we apply another round of scaling. This is done using StandardScaler from scikit-learn \cite{scikit-learn}.

This method of creating a vector is based on the use of Laplacian Eigenvalues \cite{delara2018simple}. This method first transforms the adjacency matrix into its Laplacian - another form of graph matrix representation. The eigenvalues are taken of this and ordered. We found that the use of a vector created this way and one created with the more straightforward method described above gives similar results. 

We prepare 1000 background samples and 200 signal samples for use. The background set is then split into 800 training examples and 200 test examples. The number of events we use is limited due to the nature of sampling from a simulated GBS device, which is discussed in Section \ref{Sec:photonic}. 

K-means clustering is a popular method of unsupervised classification \cite{1056489, Celebi_2013}. Its aim is to separate samples into several clusters. Each cluster has a centroid that is defined by the mean $\mu$ of the samples inside it. To begin finding these clusters the centroids can be initialised randomly. Then, they are updated by the algorithm by repeating 2 steps: (i) assign every point a label. This label describes what cluster it belongs to and is decided by the centroid closest to the point. (ii) The positions of the centroids are then updated by taking the average of every point in the cluster. These steps repeat until the change in centroid location is less than a selected value. As the centroids are updated the process aims to minimise the within-cluster sum-of-squares 
\begin{equation}
\label{eq:wcss}
\sum_{i=0}^{n}  \sum_{x \in C_i}  \left \| x - \mu_i \right \|^2,
\end{equation}
where $C$ is the set of clusters $(C_1, ..., C_n)$ each with a mean $(\mu_1, ..., \mu_n)$. In our method, we only cluster our samples into one cluster.  In this scenario, clustering is a somewhat-trivial problem. However, our choice of algorithm here is simply to develop a baseline to allow us to compare our classical sampling technique to a quantum equivalent. Regardless, the cluster is created using the 800 background training samples and results in the centroid $c$. When testing on a sample, $x$, we can then define a loss function 
\begin{equation}
\label{eq:loss_k}
L = \mbox{distance}(c, x), 
\end{equation}
which measures the distance between a point and the cluster centroid. Intuitively, it can be assumed points closer to the centroid will more likely come from the background sample while points further away will more likely be signal. Therefore, a point with a larger loss value can be classified as an outlier and, thus, as signal. 

\begin{figure}[!t]
	\centering
	\begin{subfigure}{0.49\linewidth} \centering
		\includegraphics[width=\textwidth]{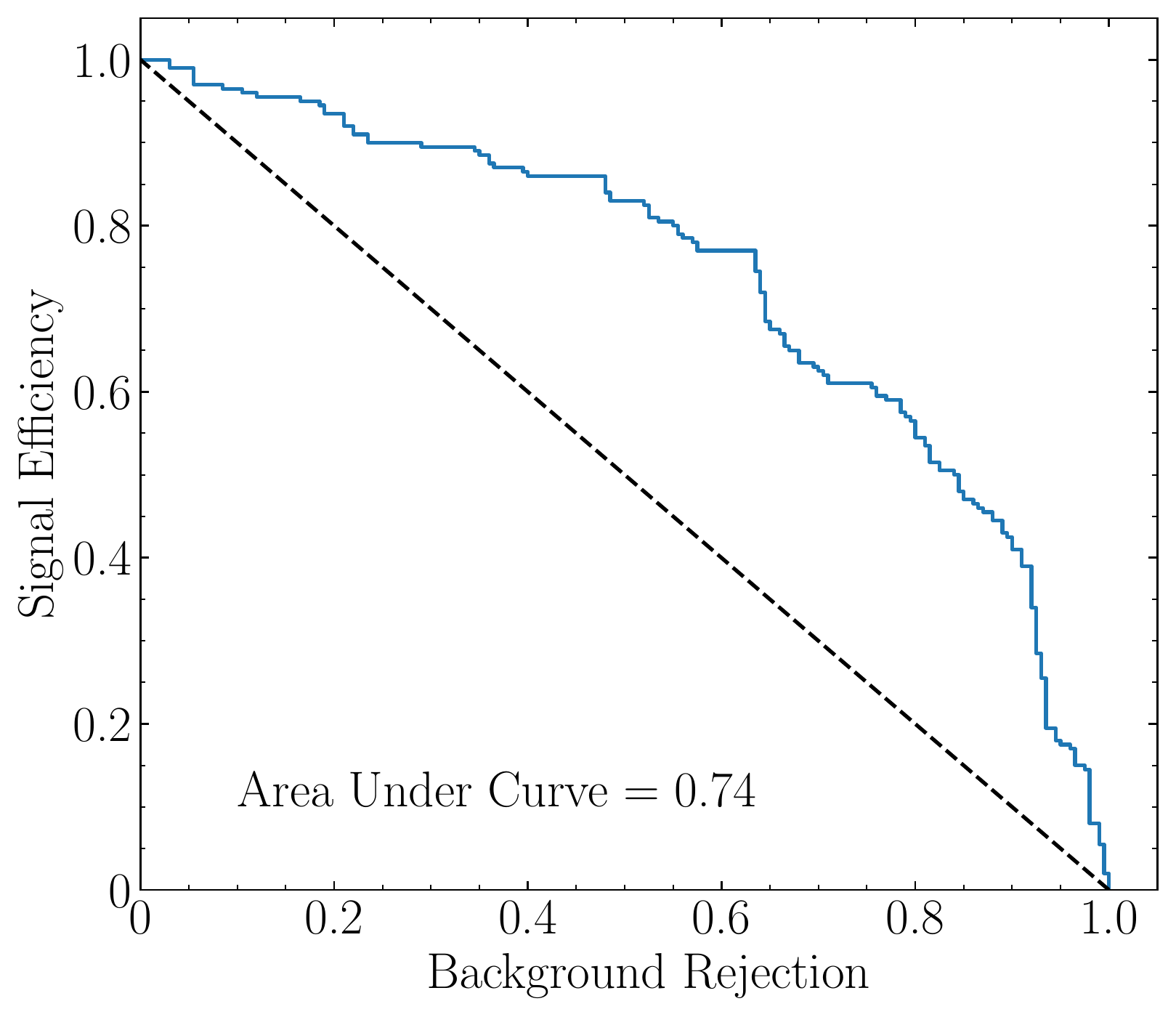}
		\caption{}
	\end{subfigure}
	\begin{subfigure}{0.49\linewidth} \centering
		\includegraphics[width=\textwidth]{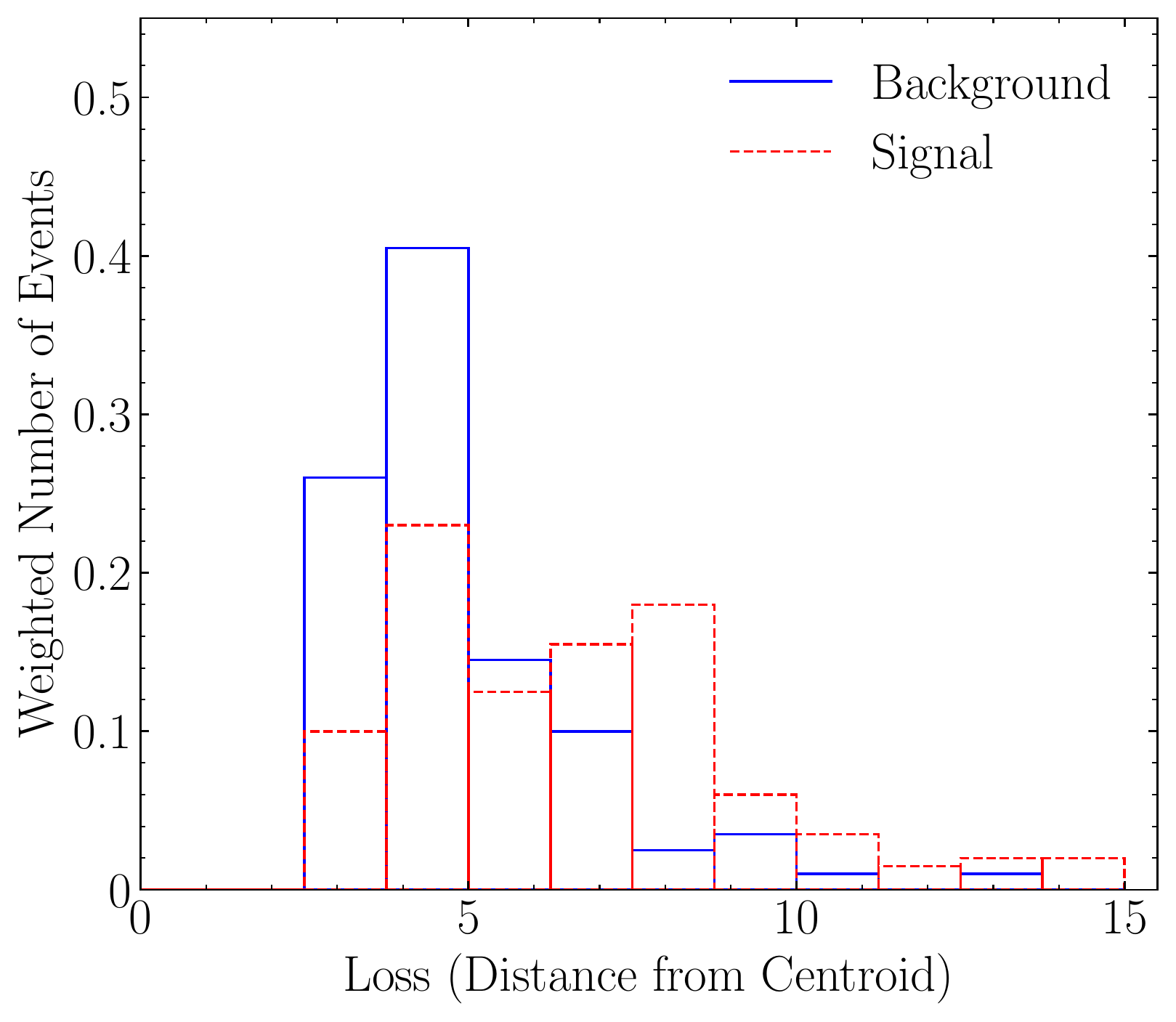}
		\caption{}
	\end{subfigure}
	\caption{a) shows the ROC curve for K-means anomaly detection method trained on classically constructed feature vectors. The distribution of the loss for the signal and background test sets is shown in b)}
	\label{fig:classical_kmeans}
\end{figure}

Figure \ref{fig:classical_kmeans} (a) shows the ROC curve from testing the fitted K-means algorithm, giving a result of 0.74 AUC. Figure \ref{fig:classical_kmeans} (b) shows the distribution of distances for the background and signal test sets.  

To give some context for this result we can compare it to another classical method of anomaly detection. This will be done using an autoencoder. Autoencoders are neural networks where the input and out dimensions match while the middle of the network forms a bottleneck. The network's optimisation task is to recreate the input exactly in the output, but this, of course, is impossible due to the smaller number of nodes in the centre of the network. Like the K-means method, the autoencoder is only trained on background data. By comparing the loss associated with background test data and signal data one can create an anomaly detector \cite{Blance:2019ibf}. We find this method gives the same AUC of 0.74.

 \section{\label{Sec:photonic}Anomaly Detection on a Photonic Quantum Devices}

In Section \ref{Sec:classical} we described a classical method of embedding the matrices we created into a vector and classifying them. Here, we replace the embedding procedure with an equivalent that can be performed on a quantum device. 

The continuous-variable quantum computing regime differs from traditional, discrete, qubit-based quantum computing in many key areas. In the qubit system we can describe states as being in the form 
\begin{equation}
\label{eq:qubit_sys2}
|\psi\rangle = c_0 |0\rangle + c_1 |1\rangle .
\end{equation}
However, when we move to the CV form of quantum computing  our states are no longer represented as qubits, but rather as qumodes. Qumodes are vectors expressed in an infinite dimensional uncountable basis
\begin{equation}
\label{eq:cv_sys2}
|\psi\rangle =\int \psi(x)  |x\rangle dx.
\end{equation}
A specific qumode $k$, at its simplest, can be described as a harmonic oscillator \cite{Braunstein_2005}
\begin{equation}
\label{eq:harm_osc}
\hat{H}_k = \frac{1}{2}(\hat{p}^2_k + \omega^2_k\hat{x}^2_k),
\end{equation}
 with the position and momentum operators 
\begin{align}
\hat{p}_k &= -\sqrt{\frac{\hbar\omega_k}{2}} (\hat{a}_k - \hat{a}^\dagger_k), \\
\hat{x}_k &= \sqrt{\frac{\hbar}{2\omega_k}} (\hat{a}_k + \hat{a}^\dagger_k).
\end{align}
These operators both depend on the creation and annihilation ladder operators $\hat{a}$ and $\hat{a}^\dagger$. By using multiple modes, and evolving their state, we can build a continuous variable quantum computer.  

While having an infinite dimensional basis is a fundamental change of the configuration space of our quantum system compared to Eq.~(\ref{eq:qubit_sys2}), the dynamics of the state $|\psi\rangle$ are still governed by unitary operators. Thus, by embedding data into a qumode one can apply gates to evolve the state, which is then measured at the end to obtain the result. However, the gate-set available to construct these circuits differ between discrete quantum-gate computers and CV devices \cite{Braunstein_2005}. 

A gate can take the form of the unitary $U = \exp(-itH)$, where H is the generating Hamiltonian with time t. On a CV device we can classify the gates into two groups based on the form of this generator. Gaussian gates are at most quadratic, and take one or two modes as input. Squeezed states can be created using the squeeze gate
\begin{equation}
\label{eq:squeeze}
S(z) = \mbox{exp}\left(\frac{1}{2}(z^*\hat{a}^2 - z\hat{a}^{\dagger^2})\right),
\end{equation}
where $a$ and $\hat{a}$  are the ladder operators and $z=re^{i\phi}$. 
The value $r$ controls the squeeze amount and $\phi$ is the squeeze phase angle.  The gate, applied to a mode, performs the transformation
\begin{align}
\label{eq:squeeze_explain}
\hat{S}^\dagger(z) \hat{a} \hat{S}(z)  &= \hat{a}\mbox{cosh}(r) +  \hat{a}^\dagger e^{i\phi} \mbox{sinh}(r), \\
\hat{S}^\dagger(z) \hat{a}^\dagger \hat{S}(z) &= \hat{a}^\dagger\mbox{cosh}(r) +  \hat{a}e^{-i\phi} \mbox{sinh}(r).
\end{align}
For a position-momentum pair, the squeezing gate will amplify one while reducing the other. States can be rotated through the rotation gate
\begin{equation}
\label{eq:rotation}
R(\theta) = \exp \left(i\theta \hat{a}^\dagger \hat{a} \right).
\end{equation}
Here, $\theta$ controls the rotation angle. The gates application on a state will rotate a states position and momentum 
\begin{align}
\label{eq:rot_expl}
R^\dagger(\phi)\hat{x}R(\phi) &= \hat{x}\mbox{cos}(\phi) - \hat{p}\mbox{sin}(\phi), \\
R^\dagger(\phi)\hat{p}R(\phi) &= \hat{p}\mbox{cos}(\phi) - \hat{x}\mbox{sin}(\phi).
\end{align}
The final Gaussian gate we will introduce is the beamsplitter gate. This a 2 mode gate (ie. it takes two qumodes as input) that transforms states $\left | \hat{a}_1, \hat{a}_2  \right \rangle$ to $\left | \hat{a}_1', \hat{a}_2'  \right \rangle $, where 
\begin{align}
\label{eq:bs_effect}
\hat{a}_1'  &= \hat{a}_1 \mbox{cos}(\theta) - \hat{a}_2 e^{-i\phi} \mbox{sin}(\theta), \\
\hat{a}_2'  &= \hat{a}_2 \mbox{cos}(\theta) + \hat{a}_1 e^{-i\phi} \mbox{sin}(\theta).
\end{align}
In this case $\phi$ and $\theta$ are parameters in the beamsplitter gate
\begin{equation}
\label{eq:beamsplitter}
B(\theta, \phi) = \mbox{exp}\left (\theta(e^{i \phi}\hat{a}_1\hat{a}_2^\dagger  - e^{-i \phi}\hat{a}_1^\dagger  \hat{a}_2) \right).
\end{equation}
As well as Gaussian gates there are also a set of non-Gaussian gates. These gates differ from Gaussian gates as they will only have 1 mode as input, and also through the degree of $H$. If the hamiltonian in the unitary has as degree of 3 or more (the position, momentum or a ladder operator in the gate is cubed, or more) it is classed as non-Gaussian. An example of this is the cubic phase gate
\begin{equation}
\label{eq:cubic_phase}
V(\gamma) = \exp \left(i \frac{\gamma}{6}\hat{x}^3 \right)
\end{equation}
With a set of Gaussian gates it is possible to construct all quadratic unitaries. However, a gate-set combining both Gaussian and non-Gaussian gates allows for universal quantum computation. This is defined by the computers ability to implement in a finite number of steps, with arbitrary precision, a unitary which is polynomial \cite{Lloyd_1999}. 

Continuous-variable photonic quantum devices have been shown to solve some classically difficult problems. One such problem is Gaussian boson sampling. This is the quantum method we will use to embed our matrices. For now, the classification step (K-means) will remain unchanged. 

\subsection{\label{Sec:GBS}Gaussian Boson Sampling}

An area where photonic devices are proposed to demonstrate an advantage is through boson sampling \cite{Hamilton_2017, aaronson2010computational}. Boson samplers are made up of an array of emitters designed to emit single photons into an interferometer. Samples are found from the regularity of photons seen by the detectors after being outputted from the device. By embedding matrices into the device one can use the device to sample the object. To create samples of our graphs from a quantum device we will embed the graph adjacency matrices into the GBS device. From the samples that will be output one can create a feature vector to use as classifier input \cite{Bromley_2020, Br_dler_2018, Schuld_2020}.

A simple analogue to boson sampling is the Galton board. A Galton board is built from a vertical board with a series of pegs attached. Balls are dropped into the device and make there way down the board, their path being altered by the pegs they hit. At the bottom of the board the balls are collected. In the boson sampling device however bosons can be emitted from multiple locations, instead of just one. Similar to the balls in the Galton board, the bosons do not interact with each other, To encode information into the device one can change the layout of the ``pegs". The probability of finding a ball in a specific bin (or a photon at a specific detector) requires the calculation of the permanent

\begin{equation}
\label{eq:perm}
\mbox{Per}(A) = \sum_{\pi \in S_N} \prod^n_{i=1} A_{i, \pi(i)}.
\end{equation}

The permanent is found from the matrix A, whose entries $A_{ij}$ will be the probability that a photon $i$ will be found in detector $j$, and $S_N$, the entire set of permutations of N photons. The permanent can also be defined by its relation to graphs. Calculating the permanent of the adjacency matrix of a bipartite graph is the equivalent of summing all the graphs perfect matchings. A graph, or subgraph, can be described as a ``matching" if every vertex in the graph is connected with, at most, one other vertex. A ``perfect matching" is the situation when every vertex is connected to one other vertex.  The calculation of the permanent is a $\#$P-problem and hard to solve on classical computers \cite{VALIANT1979189}.

Boson Sampling requires the generation of deterministic sources of single photons - something not currently available. Extensions to the above boson sampling model have been proposed to get around this. One of these is Gaussian boson sampling (GBS) \cite{Hamilton_2017}. GBS, instead of single photon states, uses single mode squeezed states. This setup has the same problem complexity as typical boson sampling. However, since squeezed states can be generated deterministically, it does not have the same scaling issues.

Unlike ``standard" boson sampling, which requires the calculation of the permanent, to simulate Gaussian boson sampling requires the calculation of the hafnian. The hafnian and permanent are related through
\begin{equation}
\label{eq:perm2haf}
\mbox{Perm}(M) = \mbox{Haf}\begin{pmatrix}
0 & M\\ 
M^t & 0
\end{pmatrix}
\end{equation}
where M is a matrix. The hafnian will calculate how many perfect matchings are in an arbitrary graph. If the edges are weighted (ie. there is a weighted adjacency matrix) the hafnian calculation will sum the weights associated with the vertices of the perfect matchings. The hafnian is calculated as
\begin{equation}
\label{eq:hafnian}
\mbox{Haf}(M) = \sum_{\pi \in P_n} \prod_{(u,v) \in \pi} M_{(u,v)},
\end{equation}
where $M$ is a generic symmetric adjacency matrix. Here, $P_n$ is the set of all permutations of pairs from a list of indices $n$ long. If $n=4$, then $ P_n = [\{(1,2),(3,4)\}, $ $\{(1,3),(2,4)\}, \{(1,4), (2,3)\}]$. For a graph containing $n$ nodes, $P_n$ would find every set of perfect matchings. Note, in the $n=4$ example that the three sets found are the indices of nodes in the three possible perfect matching sets. If $n$ were to equal an odd number, there could not be a perfect matching and the hafnian would equal zero. The hafnian is a more general function than the permanent but like it, there is no known method of classically calculating it efficiently. 

The state prepared by a GBS device with $N$ modes is fully described by a $2N\times 2N$ covariant matrix $\sigma$ and a displacement term, $d$. Here, we focus solely on $\sigma$. For such a state the GBS device can be sampled. What will be outputted is an array of photon counts. The array $\bar{n}$ will be filled such that $\bar{n} = [n_1, ..., n_N]$, where each $n_i$ represents how many photons have been counted at each detector.  The probability of seeing a result $\bar{n}$ is
\begin{equation}
\label{eq:GBS_dist_0}
P(\bar{n}) = \frac{1}{\sqrt{\mbox{det}  (Q )}}\frac{  \mbox{Haf}(\tilde{A}_{\bar{n}}) }{\bar{n}!}.
\end{equation}
Here, $\bar{n}!=n_1!n_2!... n_N!$. The matrix $\tilde{A}_{\bar{n}}$ and Q are both related to the covariance matrix $\sigma$. This can be seen through the relations $Q = \sigma + I/2$, $\tilde{A} = X(I - Q^{-1}) $ and $X=\bigl(\begin{smallmatrix}
0 & I \\ 
I& 0
\end{smallmatrix}\bigr)$ \cite{Hamilton_2017}. In Eq.~(\ref{eq:GBS_dist_0}) the hafnian of $\tilde{A}$  is not being found, but rather a modified version $\tilde{A}_{\bar{n}}$ . The new matrix will be the same, except for the removal (or duplication) or certain rows and columns from $\tilde{A}$, depending on the values in $\bar{n}$. If  $\bar{n}_i = 0$, then the i-th row and column of $\tilde{A}$ will not be included, if $\bar{n}_i = 1$ the row/ column will not change, and if $\bar{n}_i > 1$ the row/ column will be duplicated. Finally, if all values in $\bar{n}$ are $1$ then $\tilde{A}=\tilde{A}_{\bar{n}}$. 

For our purposes, we need to be able to embed our graphs into the device to retrieve samples. Therefore, there is a need to relate the adjacency matrices $A$ to the covariance matrix $\sigma$. This is done through the double encoding strategy
\begin{equation}
\label{eq:doubled}
\tilde{A} = A \oplus A.
\end{equation}
Eq (\ref{eq:GBS_dist_0}) can be written such to show the probability of sampling a photon event $\bar{n}$ with respect to our adjacency matrices
\begin{equation}
\label{eq:GBS_dist}
P(\bar{n}) = \frac{1}{\sqrt{\mbox{det} (Q )}}\frac{ \left | \mbox{Haf}(A_{\bar{n}}) \right |^2}{\bar{n}!}.
\end{equation}
Again, the hafnian of the entire matrix $A$ is not found, but rather a submatrix $A_{\bar{n}}$. Practically, to sample from this distribution we embed the matrix into the device via parameters in the squeezing gates and interferometer of our photonic circuit.

Overall, the probability of a set of photons $\bar{n}$ being detected is therefore proportional to the weighted number of perfect matchings of the subgraph $A_{\bar{n}}$. By sampling from a GBS device we can create feature vectors to use as input to classifiers.  

\subsection{\label{Sec:GBS_device}Constructing a GBS Circuit}

\begin{figure}[!t]
	\centering
	\begin{subfigure}{0.99\linewidth} \centering
		\includegraphics[width=\textwidth]{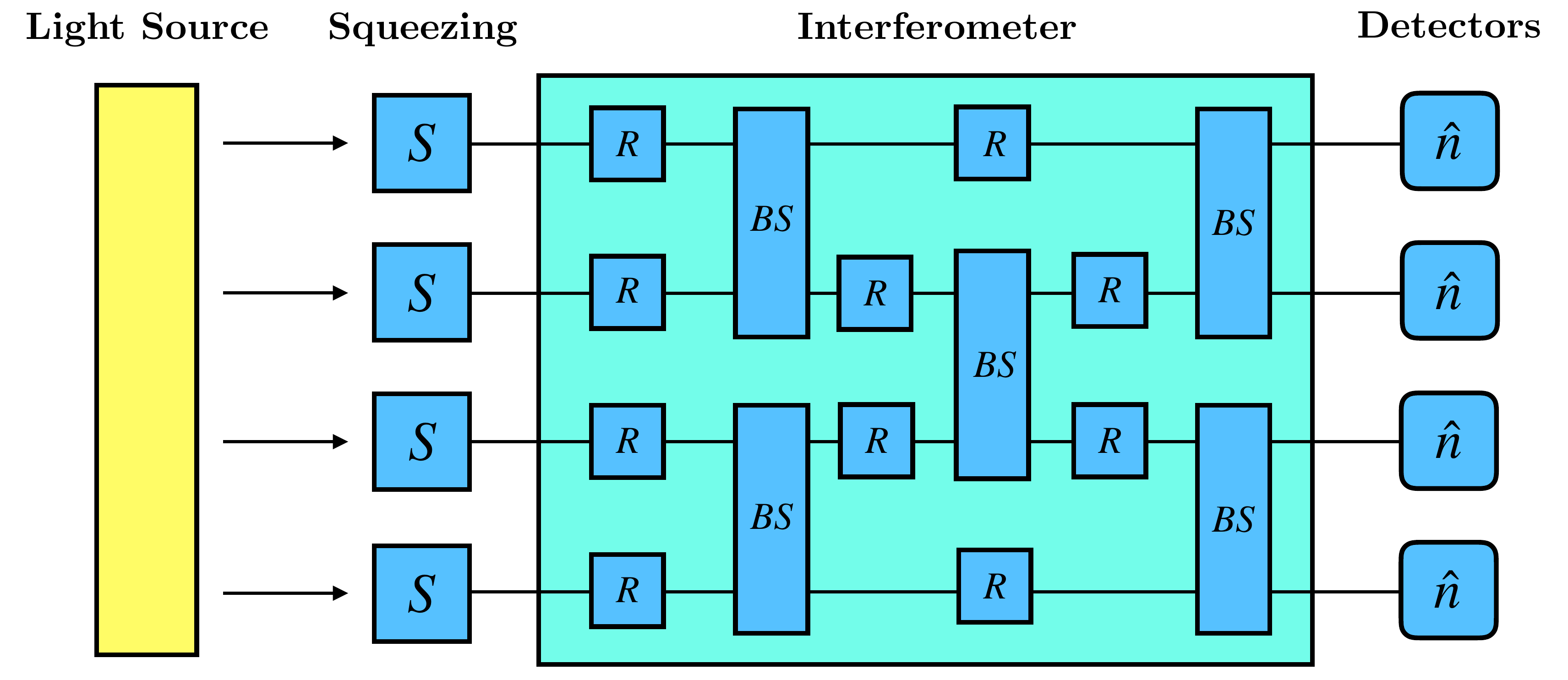}
	\end{subfigure}
	\caption{A Gaussian boson sampling device. It consists of a set of squeezing gates, an interferometer and photon detectors. The interferometer itself is constructed from rotation gates and beamsplitters.}
	\label{fig:gbs_device}
\end{figure}

Due to the hafnian calculation, simulating Gaussian boson sampling is difficult on a classical machine. However, using quantum gates one can build a Gaussian boson sampling device that will run on a photonic quantum computer. The sampler is created from squeezing, rotation and beamsplitter gates and is shown in Figure \ref{fig:gbs_device}. The gate parameters (detailed in Eqs.~(\ref{eq:squeeze}), (\ref{eq:rotation}) and (\ref{eq:beamsplitter})) are all determined by the adjacency matrix $A$. The matrix $A$ can be decomposed, using the Takagi-Autonne decomposition, to obtain
\begin{equation}
\label{eq:takagi}
A = U \mbox{diag}(\lambda_1, ..., \lambda_N)U_T.
\end{equation}
Here, U is a unitary matrix and the set of $\lambda$'s are the matrix eigenvalues \cite{Bromley_2020}. It is possible to further decompose the matrix $U$ to give parameters for the set of interferometer gates \cite{Clements:16}. To define a term to use for our squeeze gates parameters we need to introduce a scaling term c. This term is defined such that
\begin{equation}
\label{eq:mean_n_c}
\bar{n} = \sum_{i=0}^{N}\frac{c\lambda_i^{2}}{1 - c\lambda_i^{2}},
\end{equation}
where $\bar{n}$ is again the mean number of photons, a parameter we can choose  \cite{Bromley_2020}. From this and the eigenvalue we can find a squeezing parameter $z_i= \mbox{tanh}^{-1}(c\lambda_i)$. Altogether, this will allow us to embed our adjacency matrices within a GBS device and produce samples from it.

The construction of the circuit on a real device allows large amounts of sampling to be performed very quickly. Current devices find around  $10^{5}$ samples every second \cite{Schuld_2020}. However, access to these devices is currently limited, resulting in the process needing to be simulated. This is accomplished by using the Python library Strawberry Fields \cite{Killoran_2019}.

\subsection{\label{Sec:GBS_vecs}Creating Feature Vectors from GBS Samples}

By first transforming graphs into their adjacency matrix representation (as carried out in Section \ref{Sec:data}) they can then be embedded into the GBS device. When the device is sampled the process will therefore be driven by the information contained in the matrix. The generated samples can be used to create feature vectors to use in a classification procedure. The output from the GBS device will be arrays $N$ long, where $N$ equals the number of nodes in our graph. This vector will be made from information on frequencies of measurements from the device. As we run the device, we can choose from either thresholding the output, or not. In ``threshold" mode, the output sample of the device only tracks if a photon has been detected in a specific mode or not, rather than count how many have been seen. On a simulator, not thresholding the results proved very time-intensive. Therefore, the sample results where thresholded. The output of the device will therefore be an array $N$ long, filled with zeroes or ones depending on whether a photon was detected. 

Regardless, sampling using the GBS simulator is time-consuming with or without photon thresholds. This constrains us with the amount of data we can use. As mentioned, feature vectors are built from frequencies of certain measurements from within an events set of samples. It follows that we need to sample each graph multiple times to accurately determine these measurements. The amount of samples taken from each event is another factor that influences runtime. We choose to sample each of the 5 matrices in a single event 6500 times, giving a total of 32,500 samples per event. This value gives us a good result.

The generated set of samples can then be used to calculate feature vectors. Here, a few methods exist to construct these \cite{Schuld_2020}. An option is to craft a vector that contains probabilities of a photon being seen at each detector. If we have an event with a device containing $k$ photon detectors we can construct a vector

\begin{equation}
\label{eq:gbs_feat_vec}
v = (p_{k_{1}}, p_{k_{2}}, ..., p_{k_{N}}).
\end{equation}

Here, $p_k$ is the chance a photon will be seen in this position by the detector. This setup, though it can be computationally expensive at large numbers of photons, grants us a lot of freedom to create the vector. 

For each graph, we have a set of five adjacency matrices. For each matrix a feature vector is found (just as in Eq.~(\ref{eq:gbs_feat_vec})). The five vectors representing the graph are then combined, making a vector of length 30. This procedure mirrors how we handled the classical eigenvalue scenario. However, here we create the samples from the GBS device.

\subsection{\label{Sec:point_processing}K-means Anomaly Detection using Gaussian Boson Sampling}

To set a benchmark for anomaly detection with samples from the GBS device we will once again use K-means clustering. The pattern for this will match what was done in Section \ref{Sec:classical}. The vectors will, firstly, be scaled using scikit-learn's StandardScaler. The K-means algorithm will then be run on 800 background samples to find a centroid. Distances can be calculated from the centroid to points in our test set of 400 background and signal graphs. This distance will be used as an error value in the fit, as shown in Eq.~(\ref{eq:loss_k}).

\begin{figure}[!t]
	\centering
	\begin{subfigure}{0.49\linewidth} \centering
		\includegraphics[width=\textwidth]{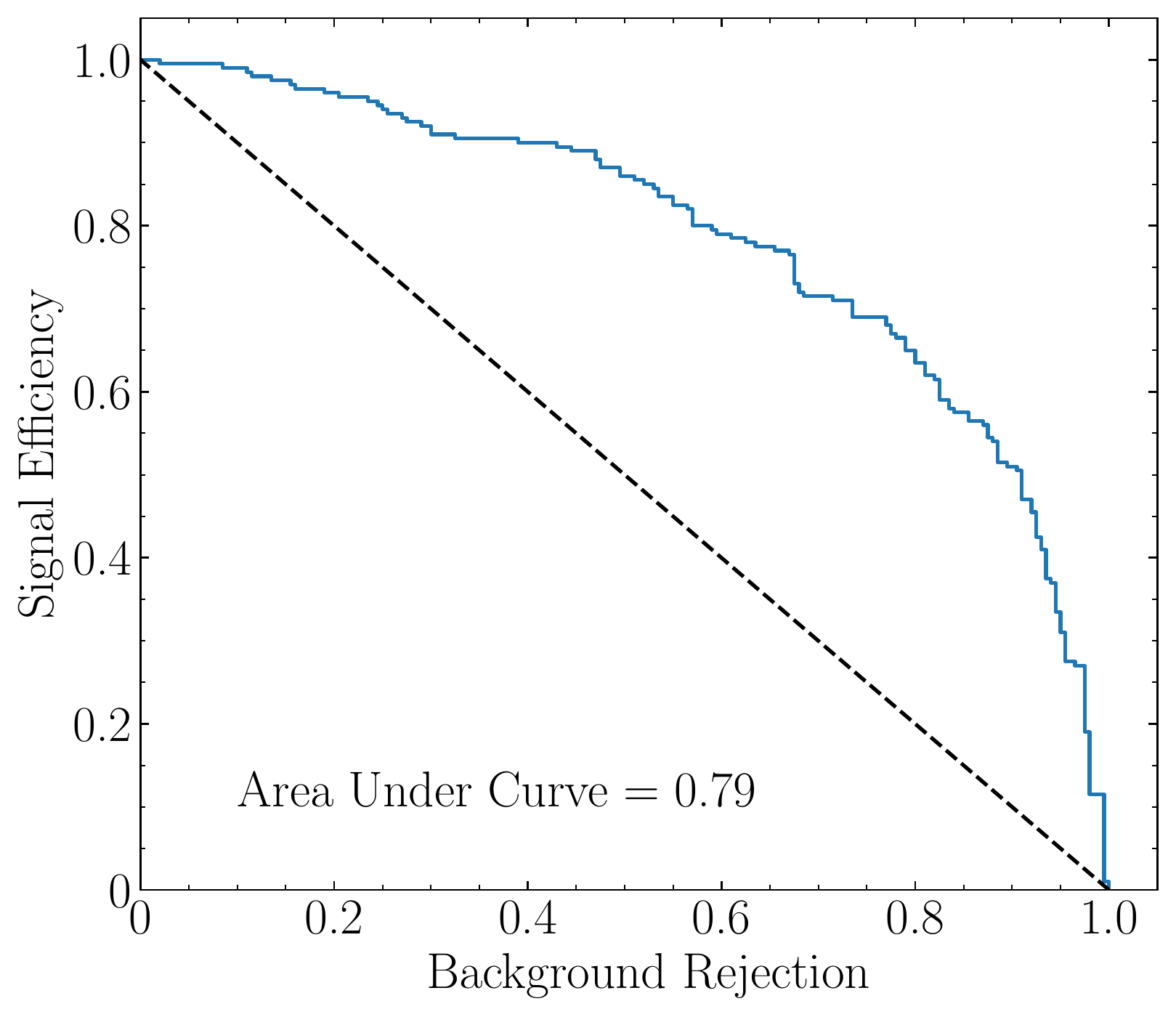}
		\caption{}
	\end{subfigure}
	\begin{subfigure}{0.49\linewidth} \centering
		\includegraphics[width=\textwidth]{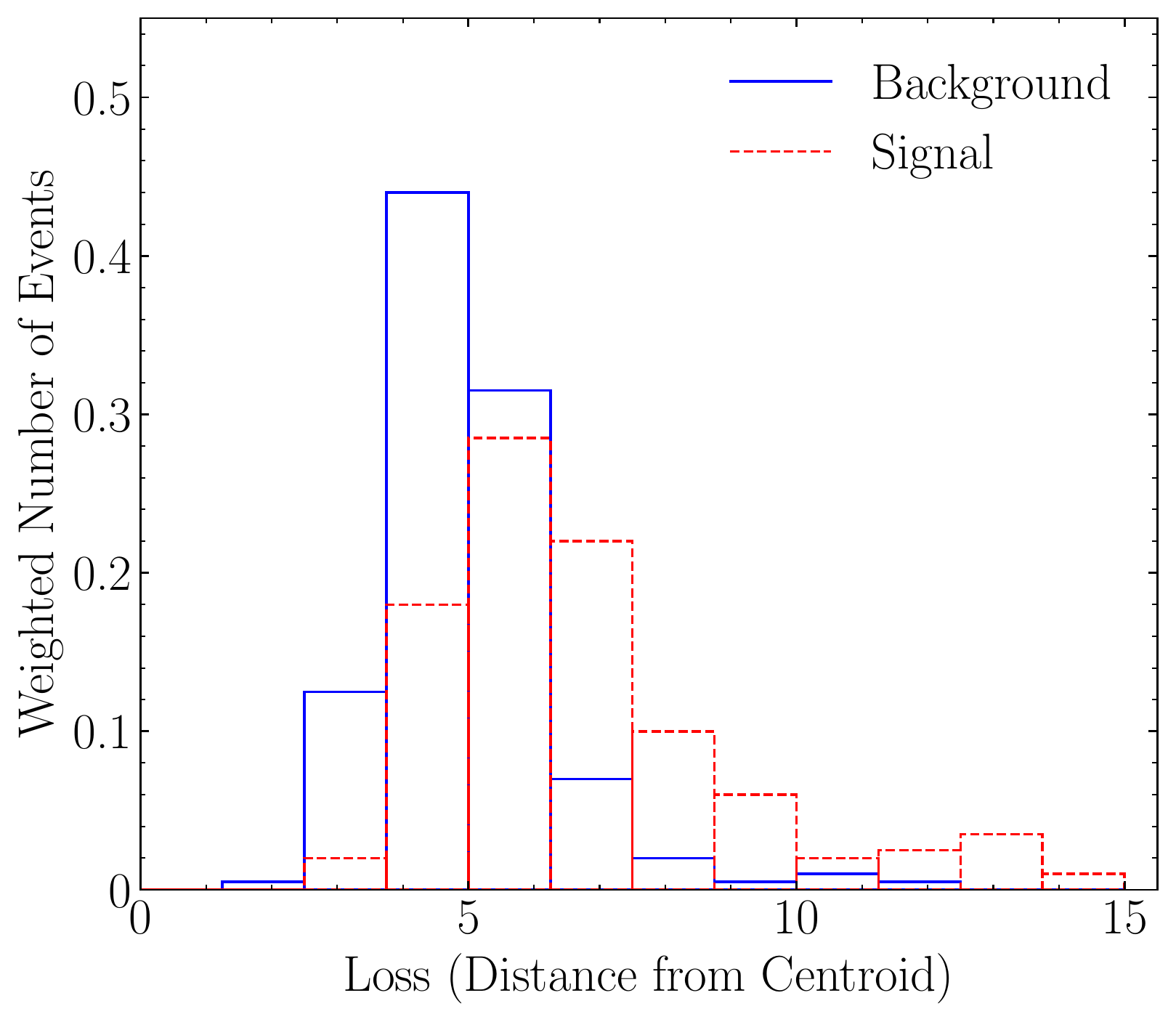}
		\caption{}
	\end{subfigure}
	\caption{a) shows the ROC curve for K-means anomaly detection method trained on feature vectors constructed from GBS samples. The distribution of the loss for the signal and background test sets is shown in b)}
	\label{fig:gbs_kmeans}
\end{figure}

Results for this method are shown in Figure \ref{fig:gbs_kmeans}. We achieve an AUC score of 0.79. This is an improvement over the classical scenario where an AUC of $0.74$ was achieved. The GBS sampling method appears to provide an advantage in creating samples to use for anomaly detection tasks. 

\section{\label{Sec:qmeans}Q-means Clustering}

As mentioned in Section \ref{Sec:Intro} we can consider our anomaly detection problem in three parts: (i) graph creation, (ii) embedding of matrices and (iii) the classification method. In Section \ref{Sec:GBS} we discussed a quantum method to embed the matrices. In this section, we introduce Q-means clustering, a quantum equivalent of K-means clustering, which we will use for anomaly detection.

K-means can be described as having two steps: (i) assigning points a label and (ii) updating the centroids. To successfully complete part (i) there is a need to be able to deduce what centroid each point is closest to. In this variant of Q-means we focus specifically on this.  

As with the GBS sampling device, the Q-means method is applied using a quantum circuit. The circuit constructed aims to provide a measure of how close a point is to a centroid. The circuit itself can be split into three parts: data embedding, distance calculation and readout. After calculating which centroid the point is closest to we can then move onto the second step of the clustering procedure, updating the centroids. The centroids are chosen such that they are the mean of every point in the cluster. These steps will be repeated until the distance between the updated cluster location and the previous location is less than a threshold $\epsilon$ \cite{lloyd2013quantum}.

We will first view the task of Q-means classification in the discrete qubit scenario. To embed our data to use in the circuit we will use $U_3$ gates,
\begin{equation}
\label{eq:u3}
U_3(\theta, \phi, \lambda) = \begin{bmatrix}
\cos(\frac{\theta}{2}) & -\exp(i\lambda)\sin(\frac{\theta}{2})\\ 
\exp(i\phi)\sin(\frac{\theta}{2}) & \exp(i(\phi+\lambda))\cos(\frac{\theta}{2})
\end{bmatrix}.
\end{equation}
The vectors are embedded in the $\theta$, $\phi$ and $\lambda$ parameters. This takes the values and embeds them as angles in a qubit. For vectors with more than three features we must use more than one qubit.

After being embedded the states must enter into a circuit designed to measure some metric of distance. This will be done using the SwapTest circuit. See Appendix \ref{appendix:Qmeans_O} for more information. For two states $\left | \alpha \right \rangle$ and  $\left | \beta \right \rangle$ the SwapTest routine will measure the overlap between them. If we find a probability of $P(\left | 0 \right \rangle)=0.5$ then the two states are orthogonal, if $P(\left | 0 \right \rangle)=1$ then the states are the same. If the centroids are embedded into a state alongside a sample vector we can measure which centroid state overlaps the most with the vector. The centroid with the most overlap will be chosen as the closest and the vector will be assigned to that cluster. 

\begin{figure}[!t]
	\centering
	\begin{subfigure}{0.89\linewidth} \centering
		\includegraphics[width=\textwidth]{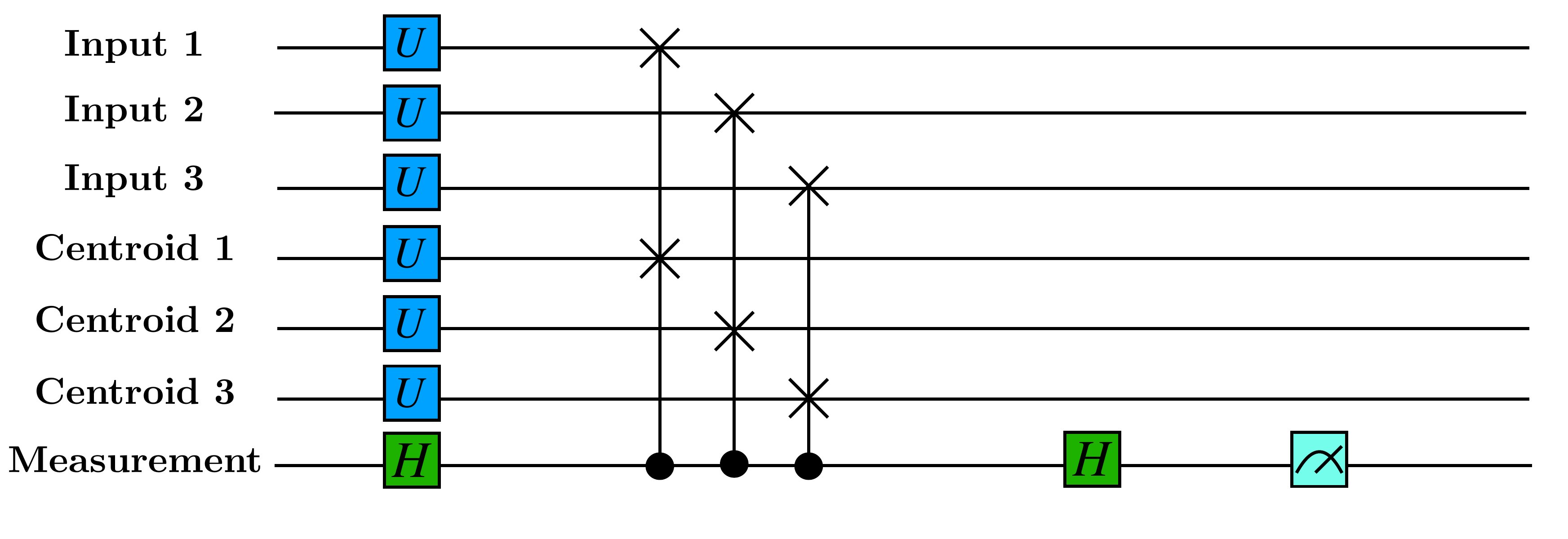}
	\end{subfigure}
	\caption{Diagram of the Q-means quantum circuit. The feature vector and centroid is encoded through $U_3$ gates, while a SwapTest is used to provide a measure of their similarity.}
	\label{fig:qmeans}
\end{figure}

SwapTest circuits are constructed from Hadamard $H$ and SWAP gates, respectively
\begin{equation}
\label{eq:H}
H = \frac{1}{\sqrt{2}}\begin{bmatrix}
1 & 1\\ 
1& -1
\end{bmatrix}
\end{equation}
and
\begin{equation}
\label{eq:swap}
SWAP = \begin{bmatrix}
1 & 0 & 0 & 0\\ 
0 & 0 & 1 & 0\\ 
0 & 1 & 1 & 0 \\ 
0& 0 & 0 & 1
\end{bmatrix}.
\end{equation}
A CSWAP gate is similar to a SWAP gate, Eq.~(\ref{eq:swap}), but depends on the state of a control bit. The Hadamard gate is used regularly when building quantum circuits to introduce entanglement into the system.
These gates, alongside the $U_3$ embedding gates are combined to form the circuit shown in Figure \ref{fig:qmeans}. The circuit shown is designed for a model where vectors contain between 6 and 9 features, and hence need to use 3 qubits to store the information. The vectors created from the GBS method contain 30 elements and hence will require 10 qubits. 

We have focused on a single variant of Q-means here. However, the model can be expanded to improve its performance further. Including other quantum algorithms (such as Grovers) can improve the label assigning stage \cite{lloyd2013quantum}, while a deeper reliance on qRAM can allow for an improved scaling complexity with respect to the data-set size itself \cite{kerenidis2018qmeans}.

What is appealing about the construction of this circuit is that (assuming states are prepared) the SwapTest routine does not depend on the number of points in a vector. To encode a state into quantum memory has the time complexity of $\mathcal{O}(\mbox{log}(N))$, where $N$ is the number of features in our vector \cite{lloyd2013quantum, kopczyk2018quantum}. This matches with what we may intuitively expect - for $N$ features we require $\log_2(N)$ qubits. For regular K-means we expect a time complexity of $\mathcal{O}(N)$. By using a quantum variant of K-means we see an exponential performance increase, with respect to the number of features in our sample.

\subsection{Implementation, Performance and Scalability}

The discussion in the previous section is based on the discrete qubit quantum computing regime. However, this methodology is also possible to carry out in the continuous-variable, qumode scenario. The Q-means method presented here is based on using the SwapTest to measure the overlap of two states. This has been shown to be the equivalent of Hong-Ou-Mandel effect \cite{Garcia_Escartin_2013}. The Hong-Ou-Mandel effect is a way to compare two arbitrary states in quantum photonics. Making use of this would allow one to implement Q-means on a photonic device straightforwardly. 

To implement the Q-means quantum circuit shown in Figure \ref{fig:qmeans} we use the Python package PennyLane \cite{bergholm2020pennylane}. 
The Q-means algorithm is trained with the same parameters as the previous two scenarios - using 800 background samples, each with 30 features. We use both methods of generating samples to train the model. In both cases, the classical eigenvalue scenario and the GBS sample variant, we find Q-means and K-means results to be equivalent. If one judges the Q-means performance based on the potential efficiency improvements there is a clear benefit to its use. The power of Q-means lies in its ability to scale to larger feature-vectors. The increase in the availability of quantum devices, and the ability to run these algorithms on their native devices, allows clustering methods to be more viable.

\section{\label{Sec:conc}Conclusions}

Separating rare unspecified signal events from common Standard Model backgrounds is one of the most important tasks multi-purpose experiments at the LHC perform. Data-driven searches, ones that train directly on the data without making assumptions about any features in new physics models, are an appealing option. By learning solely the distributions of the background Standard Model events it is possible to flag events that do not share similar features as signal. When building models and constructing datasets to train on it is worth considering how data is represented. Particle physics events are well-suited to be stored as graph structures, allowing relationships between event constituents to be naturally captured. To use these graph structures for classification one needs to use a model the structure naturally fits (such as a graph neural network) or be able to embed the structure (for example, into a kernel or vector). 

Continuous-variable (CV) quantum computing, through the use of Gaussian Boson Sampling, provides a method of embedding a graph into a lower-dimensional representation. CV quantum computers differ from discrete, qubit, models of quantum computing by their reliance on infinite-dimensional qumodes. These devices offer unique quantum advantages, one of which is boson sampling. Boson sampling creates samples from detecting photons after they travel through an interferometer. Simulating this, due to the matrix operations involved, is a $\#$P problem and therefore a classically hard task. 

We propose using Gaussian Boson Sampling to embed generated data into a feature vector and perform anomaly detection with it. This method appears to be a viable choice for trigger level anomaly detection during future LHC runs because of the speed sampling can occur on a near term photonic device. Results using GBS to generate features compare favourably to results found when the feature vectors have been classically generated. Another quantum advantage can be gained from the use of Q-means clustering. We find performing anomaly detection using Q-means clustering gives equivalent results to K-means but has the potential to scale to large feature vectors more efficiently. The variant of Q-means presented here has a $\mathcal{O}(\mbox{log(N)})$ complexity, with respect to the size of the feature vector. This is compared to the K-means complexity of $\mathcal{O}(N)$. While Q-means is implemented here in the discrete qubit-based model it should be possible to expand this to the CV paradigm.

\newpage

\appendix

\section{SwapTest}
\label{appendix:Qmeans_O}

A SwapTest circuit provides a method of checking the overlap between two states. Following \cite{kopczyk2018quantum}, we can construct the circuit from Hadamard and CSWAP gates. We will use the information on the overlap of the states in our implementation of a quantum K-means routine. To understand how SwapTest operates we begin by defining a state

\begin{equation}
\label{eq:state}
\left | \psi\right \rangle = \left | 0,\alpha,\beta \right \rangle.
\end{equation}

The state $\psi$ contains the two states we wish to measure ($\alpha$ and $\beta$) which can contain multiple qubits and a control qubit. This initial state is evolved by applying a Hadamard gate to the control bit, resulting in the new state

\begin{equation}
\label{eq:state_1}
\left | \psi_1 \right \rangle =  \frac{1}{\sqrt{2}} (\left | 0,\alpha,\beta \right \rangle + \left | 1,\alpha,\beta \right \rangle ).
\end{equation}

The next gate applied is CSWAP. The controlled SWAP gate will swap qubits based on the value of the control qubit. The application of a CSWAP will therefore evolve our state to

\begin{equation}
\label{eq:state_2}
\left | \psi_2 \right \rangle  = \frac{1}{\sqrt{2}} (\left | 0,\alpha,b \right \rangle + \left | 1,b,\alpha \right \rangle ).
\end{equation}

Finally, another Hadamard gate is applied to the control qubit:

\begin{equation}
\label{eq:state_3}
\left | \psi_3 \right \rangle  =\frac{1}{2}  \left | 0 \right \rangle(\left | \alpha,\beta \right \rangle + \left | \beta,\alpha \right \rangle ) + \frac{1}{2}\left | 1 \right \rangle (\left | \alpha, \beta \right \rangle - \left | \beta,\alpha \right \rangle ).
\end{equation}

The probability of measuring the state $\left  | 0\right \rangle$ will give us a measure of the overlap. 

\begin{align}
P(\left  | 0\right \rangle)  &= \left | \frac{1}{2} \left \langle 0 | 0 \right \rangle(\left | \alpha,\beta \right \rangle + \left | \beta,\alpha \right \rangle ) + \frac{1}{2}\left \langle 0 | 1 \right \rangle (\left | \alpha,\beta \right \rangle - \left | \beta,\alpha \right \rangle )  \right |^2    \\
& = \frac{1}{4} \left | (\left | \alpha,\beta \right \rangle  +  \left | \beta,\alpha \right \rangle )  \right |^2   \\
&  = \frac{1}{4} ( \left \langle \beta | \beta \right \rangle \left \langle \alpha | \alpha \right \rangle+ \left \langle \beta | \alpha \right \rangle \left \langle \alpha | \beta \right \rangle+ \left \langle \alpha | \beta \right \rangle \left \langle \beta | \alpha \right \rangle+ \left \langle \alpha | \alpha \right \rangle \left \langle \beta | \beta \right \rangle) \\
&  = \frac{1}{2} + \frac{1}{2} \left |  \left \langle \alpha | \beta \right \rangle   \right | ^2
\end{align}

If states $\left  | \alpha \right \rangle$   and $\left  | \beta \right \rangle$ are identical then $P(\left  | 0\right \rangle)$ will equal 0, if they are orthogonal than $P(\left  | 0\right \rangle) = 0.5$

\bibliographystyle{inspire}
\bibliography{references}

\end{document}